%% file: paper.tex
\documentclass[showpacs,twocolumn,prd]{revtex4}
\usepackage{graphicx}
\usepackage{latexsym}
\usepackage{amsmath}

\newcommand{\beq}{\begin{equation}}
\newcommand{\eeq}{\end{equation}}
\newcommand{\bea}{\begin{eqnarray}}
\newcommand{\eea}{\end{eqnarray}}

\begin{document}

\title{Measuring asymmetries in flavor asymmetric machines}

\date{\today}

\author{A.A. Alves Jr.}
\affiliation{Istituto Nazionale di Fisica Nucleare - Sezione di Roma
c/o Dipartimento di Fisica - Universit\`a degli Studi di Roma "La Sapienza"
P.le Aldo Moro, 2 - 00185, Roma, Italy} 

\author{J. Magnin}
\email{jmagnin@cbpf.br}
\affiliation{Centro Brasileiro de Pesquisas F\'{\i}sicas, Rua Dr. Xavier
Sigaud 150, Urca 22290-180, Rio de Janeiro, Brazil}

\begin{abstract}
The LHC offers a unique opportunity to investigate an ample spectra of
phenomenous ranging from the Electro-Weak (EW) to the QCD sector of the Standard
Model (SM). Among the quantities which can be measured in the LHC experiments
are the CP and production asymmetries for several particles in a wide variety of
decay modes. In this work we discuss about the interplay between production and
CP asymmetries for particles produced in proton-proton interactions and the
effects of one on the measurement of the other. This kind of effects are not
present in flavor symmetric machines like the Tevatron or $e^+-e^-$ colliders.
\end{abstract}

\maketitle

\input{introduction.tex}

\input{meson.tex}

\input{cp-asymmetry.tex}

\input{conclusion.tex}

\acknowledgments{A.A. Alves Jr. is supported by the INFN fellowship
  program. A.A. Alves Jr. is grateful for the hospitality at the Physics 
  Department, Universit\`a degli Studi di Roma ``La
  Sapienza''. J.M. acknowledges very useful discussions with J. Miranda and
  C. G\"obel on the use of the method of Ref.~\cite{bediaga}.}

\input{bibliography.tex}
\end{document}

%% file: introduction.tex
\section{Introduction}

Particle-antiparticle production asymmetries  have played an important role in
the understanding of the hadronization mechanisms of partons in high energy
hadron-hadron interactions. Today there exist copious experimental
evidence~\cite{experiments} indicating that produced particles
sharing valence quarks with the initial hadrons are produced at a different
rate than particles sharing none. In fact, the so called leading particle
effect, which is responsible for the particle-antiparticle
production asymmetry, has firmly established the role of the recombination mechanism
in hadron production~\cite{asymmetry} in high energy interactions. Those studies
have also given important insights on the structure of the initial
hadrons~\cite{intrinsic-sea}.  

So far, particle-antiparticle asymmetries, both as a function of the transverse
momentum, $p_T^2$, and as a function of the scaled longitudinal momentum,
$x_F=2p_L/\sqrt{S}$, of the produced particles, have been measured in the range
of a few tenth of GeV center of mass (c.m.) energies~\cite{experiments}, and
mostly in the production of strange and charm hadrons. With the advent of LHC, it could
be interesting to measure such particle-antiparticle production asymmetries at
highest c.m. energies to understand to which extent the ratio
between the fragmentation and recombination mechanisms in the hadronization is
dependent on the c.m. energy. Furthermore, since beauty hadron production asymmetries
have not been measured at all, it could be interesting to investigate also the role of
the leading particle effects by itself in this case. However, in the case of
beauty meson production, as mesons are detected and measured through their
decay products, CP asymmetries and  mixing effects can affect the
determination of the production asymmetries, thus spoiling the study of the
hadronization mechanisms. Conversely, particle-antiparticle production
asymmetries can be an important effect, polluting weak interaction effects, in
the measurement of quantities involving the comparison of particles decays with
their charge conjugate ones in machines such LHC, which are not symmetric with
respect to particle and antiparticle production. Thus, the above mentioned
effects can be important in the determination of mixing parameters in the
$B^0_{d/s} - \overline{B}~^0_{d/s}$ system, CP asymmetries in $B^\pm$ and
$B^0_{d/s}/\overline{B}~^0_{d/s}$ decays, etc. Furthermore, the LHCb
Collaboration~\cite{lhcb} has an extensive program to measure  
$D^0-\overline{D}~^0$ mixing and possible CP violation asymmetries in the charm
sector of the Standard Model. Since those effects are expected to be small,
thought much smaller than production effects, then the interplay between the
production and CP asymmetries has to be very well understood in order to measure
the later with significative precision .

In this work we shall discuss about how the measurement of meson production
asymmetries are affected by CP asymmetries and mixing effects and viceversa, for
mesons produced in $p-p$ collisions. Along the text the discussion will be
focused in B-meson production, but most of it is directly applicable 
to D-meson production.

%% file: meson.tex
\section{B-meson production asymmetries in $p-p$ collisions}

\subsection{$B^\pm$ production in $p-p$ collisions}

In p-p collisions the production mechanisms of $B^+$ and $B^-$ mesons are
expected to be different because of the leading particle effect. In fact, since
the $B^+=(u\bar{b})$ shares a valence quark with the initial protons and the
$B^-=(\bar{u}b)$ shares none, it is expected that $B^+$s be produced at a higher
rate than $B^-$s. The above differences in the production processes of $B^\pm$
mesons in p-p collisions can be characterized by means of the so called
production asymmetry, which is defined by  
\begin{equation}
A = \frac{N_{B^+} - N_{B^-}}{N_{B^+} + N_{B^-}}\;,
\label{eq1}
\end{equation}
where $N_{B^\pm}$ are the number of $B^+$ and $B^-$ mesons produced at the
interaction point.

In order to measure the production asymmetry of Eq.~\ref{eq1}, $N_{B^\pm}$ have
to be determined by reconstructing $B^\pm$ mesons decaying into a given
final state. However, $B^\pm$ mesons could decay violating the so called CP
symmetry, thus spoiling the determination of $N_{B^+}$ and $N_{B^-}$ at the
production vertex. To be concrete, let us start by considering $B^\pm$ mesons
decaying into an arbitrary final state $f$,  
\begin{eqnarray}
B^+ &\rightarrow& f \;,
\label{eq2}
\end{eqnarray}
and its charge congugated (c.c.). If the CP symmetry is violated in the decay,
then it follows that  
\begin{equation}
A_{CP} = \frac{BR(B^+\rightarrow f)-BR(B^-\rightarrow
  \bar{f})}{BR(B^+\rightarrow f)+BR(B^-\rightarrow \bar{f})} \ne 0\;.
\label{eq3}
\end{equation}
Since
\begin{eqnarray}
N_{B^+\rightarrow f} &=& N_{B^+}\times BR(B^+ \rightarrow f)
\label{eq5a}
\nonumber \\
N_{B^-\rightarrow \bar{f}} &=& N_{B^-}\times BR(B^- \rightarrow
\bar{f})\;,
\label{eq5b}
\end{eqnarray}
then the production asymmetry of Eq.~\ref{eq1} has to be modified to
\begin{equation}
A = \frac{N_{B^+\rightarrow f} - N_{B^-\rightarrow
    \bar{f}}R}{N_{B^+\rightarrow f} + N_{B^-\rightarrow
    \bar{f}}R}\; ,
\label{eq6}
\end{equation}
where
\begin{equation}
R=\frac{BR(B^+\rightarrow f)}{BR(B^-\rightarrow \bar{f})} = \frac{1+A_{CP}}{1-A_{CP}}\;.
\label{eq7}
\end{equation}
Using Eq.~\ref{eq7}, we can rewrite Eq.~\ref{eq6} as
\begin{equation}
A=\frac{\left[\frac{N_{B^+\rightarrow f} - N_{B^-\rightarrow
    \bar{f}}}{N_{B^+\rightarrow f} + N_{B^-\rightarrow
    \bar{f}}}\right] - A_{CP}}{1 - A_{CP}\left[\frac{N_{B^+\rightarrow f} - N_{B^-\rightarrow
    \bar{f}}}{N_{B^+\rightarrow f} + N_{B^-\rightarrow
    \bar{f}}}\right]}\;,
\label{eq10}
\end{equation}
which reduces to the usual formula of Eq.~\ref{eq1} when
$A_{CP}=0$. A similar formula can be obtained for the production asymmetry as a
function of $p_T^2$ and/or $x_{\small{F}}$. The only difference
with Eq.~\ref{eq10} will be the dependence in $p_T^2$ and/or $x_F$ arising in
$N_{B^\pm\rightarrow f}$, since the $A_{CP}$ does not depend on the
momentum of the produced particle.

\begin{figure}[t]
  \includegraphics[angle=0, height=.35\textheight]{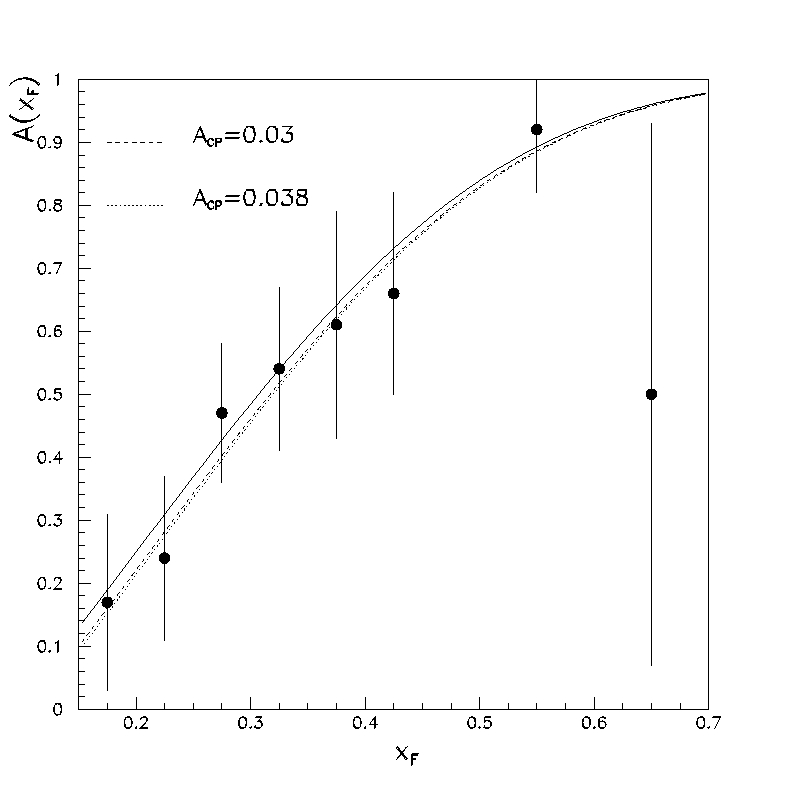}
  \caption{The $B^\pm$ asymmetry in $p-p$ collisions as a function of
    $x_F$. Full line is the fit to the SELEX $D_S^\pm$ asymmetry in
    $\Sigma^--Nucleus$ interactions given by Eq.~\ref{eq11c}. Experimental data
    has been taken from Ref.~\cite{selex}. The $B^\pm$ asymmetry is shown for
    two different values of $A_{CP}$ (See the discussion in the text).} 
\label{fig1}
\end{figure}

In order to give a numerical estimate of the effect, since there is no data
on $B^\pm$ production asymmetries in $p-p$ collisions and Monte Carlo generators
does not provide a meaningful prediction, we will assume that $B^\pm$ production
in $p-p$ interactions is similar to that of $D_s^\pm$ production in $\Sigma^- -
Nucleus$ interactions. In $\Sigma^- (dds)$-$Nucleus$ interactions, the $D_s^-$
is leading. The SELEX Collaboration~\cite{selex} has 
measured $dN/dx_{F}$ for both $D_s^-$ and $D_s^+$ and the production asymmetry
also as a function of $x_F$ in the range $[0.15,0.7]$ in 600 GeV$/c$ beam energy
interactions. The asymmetry as a function of $x_{F}$ is well represented by 
\begin{equation}
A_{D_s^\pm}(x_F) = \frac{(1-x_F)^{3.8}-1.5(1-x_F)^{7.9}}{(1-x_F)^{3.8}+1.5(1-x_F)^{7.9}}\,.
\label{eq11c}
\end{equation}
With the above assumption, the $B^\pm$ asymmetry as a function of $x_F$ and
$A_{CP}$ in $p-p$ collisions is given by
\begin{widetext}
\begin{equation}
A_{B^\pm}(x_F) = \frac{2(1-x_F)^{-4.1}-3 -
  A_{CP}\left[3+2(1-x_F)^{-4.1}\right]}{2(1-x_F)^{-4.1}+3 -
  A_{CP}\left[2(1-x_F)^{-4.1}-3\right]} \,.
\label{eq11d}
\end{equation}
\end{widetext}
In Fig.~\ref{fig1} it is shown the effect of the CP asymmetry on the production
asymmetry for $A_{CP} = 0.03\pm0.06,~ 0.038\pm0.022$~\cite{pdg} corresponding to
$B^\pm$ decaying into $\pi^+\pi^-\pi^+$ and $K^+\pi^-\pi^+$ respectively, which
are well suited modes to study $B^\pm$ production. 
The effect can be barely noted, depending however on the error in the
measurement of both, the production asymmetry and $A_{CP}$. For the
integrated production asymmetry in the range $x_F\in[0.15,0.7]$ we obtain $A=0.385\pm
0.048,~0.378\pm0.018$ respectively, while for $A_{CP}=0$ one gets $A=0.41$, which
is an effect of about $10\%$, depending on the decay mode in which the $B^\pm$s
are reconstructed.   

However, for real $B^\pm$ production in $p-p$ collisions at a c.m. energy of 7 TeV 
the production asymmetry is expected to be smaller than in interactions
at a lower c.m. energy. The leading particle effect will be still operative, but
particle production from sea-sea quark recombination is expected to be
enhanced. As this mechanism works for both, particle and antiparticle, the
production asymmetry should be smaller, thus increasing the effect of the CP
asymmetry. Once again, to have a numerical estimate let us assume that
the $B^\pm$ production asymmetry at 7 TeV c.m. energy is $A=0.2$ when measured
in a decay mode in which $A_{CP} = 0$. The raw asymmetry measured in the
$B^+\rightarrow K^+\pi^-\pi^+$ decay mode would amount to $A=0.16$, showing that
the correction by $A_{CP}$ is an effect of order $25\%$. 

\subsection{$B^0/\overline{B}~^0$ production in $p-p$ collisions}

Let us now consider $B^0/\overline{B}~^0$ production in p-p collisions. As in
$B^\pm$ production, leading particle effects are expected to
play a role since $B^0 = (d\bar{b})$ shares valence quarks with the initial
protons while $\overline{B}~^0=(\bar{d}b)$ does not. However, $B^0/\overline{B}~^0$
production is somewhat more complicated than the previous cases because of 
the $B^0-\overline{B}~^0$ mixing. 

Once again, for the sake of concreteness, let us assume $B^0\rightarrow f$ and its
c.c. decay. Once a $B^0$ is produced at the interaction vertex, it can decay
into $f$, or can oscillate to a $\overline{B}~^0$, thus decaying into $\bar{f}$. It
means that the number of measured $B^0$ and $\overline{B}~^0$ mesons decaying
into $f$ and $\bar{f}$ are related by 
\begin{eqnarray}
N_{B^0}^{exp}            &=& p_0 N_{B^0} + p_2 N_{\overline{B}~^0} \nonumber \\
N_{\overline{B}~^0}^{exp} &=& p_0 N_{\overline{B}~^0} + p_1 N_{B^0}\;,
\label{eq12}
\end{eqnarray}
where $N_{B^0},~N_{\overline{B}~^0}$ are the number of $B^0$ and $\overline{B}~^0$
mesons produced at the interaction point and
$N_{B^0}^{exp},~N_{\overline{B}~^0}^{exp}$ are the number of reconstructed $f$
and $\bar{f}$ final states, respectively. In Eq.~\ref{eq12}, $p_0,~p_1,~p_2$ are
the transition probabilities defined by~\cite{leader} 
\begin{eqnarray}
p_0 &=& \int_0^\infty{dt~P(B^0 \rightarrow B^0)} =
\int_0^\infty{dt~P(\overline{B}^0 \rightarrow \overline{B}^0)} \nonumber \\
p_1 &=& \int_0^\infty{dt~P(B^0\rightarrow \overline{B}^0)} \nonumber \\
p_2 &=& \int_0^\infty{dt~P(\overline{B}^0\rightarrow B^0)}\;.
\label{eq13}
\end{eqnarray}

Since the solution of the linear system of Eq.~\ref{eq12} is given by
\begin{eqnarray}
N_{B^0} &=& \frac{N_{B^0}^{exp} -
  (p_2/p_0)N_{\overline{B}~^0}^{exp}}{p_0-p_1p_2/p_0} \nonumber \\
N_{\overline{B}~^0} &=& \frac{N_{\overline{B}~^0}^{exp} -
  (p_1/p_0)N_{B^0}^{exp}}{p_0-p_1p_2/p_0}\;,
\label{eq14}
\end{eqnarray}
then the $B^0/\overline{B}~^0$ production asymmetry as given by Eq.~\ref{eq1} is
now 
\begin{equation}
A = 
\frac{(1 + r)N_{B^0}^{exp} - (1+\bar{r})N_{\overline{B}^0}^{exp}}
{(1-r)N_{B^0}^{exp} +  (1-\bar{r})N_{\overline{B}^0}^{exp}}\;,
\label{eq15}
\end{equation}
where  $r,~\bar{r}$ are defined as~\cite{leader}
\begin{eqnarray}
r &\equiv& \frac{p_1}{p_0} =
\left|\frac{q}{p}\right|^2\frac{x^2+y^2}{2+x^2-y^2} \nonumber \\
\bar{r} &\equiv& \frac{p_2}{p_0} = \left|\frac{p}{q}\right|^2
  \frac{x^2+y^2}{2+x^2-y^2}\;.
\label{eq16}
\end{eqnarray}
In the case of mixing with no CP violation, the production asymmetry of
Eq.~\ref{eq15} reduces to 
\begin{equation}
A = \left(\frac{1+r}{1-r}\right)
\left(\frac{N_{B^0}^{exp} - N_{\overline{B}^0}^{exp}}{N_{B^0}^{exp} +
  N_{\overline{B}^0}^{exp}}\right)\;.
\label{eq17}
\end{equation}

Notice also that no tagging needs to be used to know whether a $B^0$ or
$\overline{B}~^0$ is produced at the interaction point since the correct rate for
$B^0/\overline{B}~^0$ production is accounted for by the transition probabilities
of Eqs.~\ref{eq13}. In other words, the number of tagged $B^0/\overline{B}~^0$
mesons is given by Eqs.~\ref{eq14} in terms of the reconstructed $f$ and
$\bar{f}$ final states.

Having in mind that $x_d=0.774\pm0.008$,
$y_d^2=-0.0003\pm0.109$ and $\left|q/p\right| \sim 1$ for the $B^0_d$~\cite{pdg},
it follows that $(1+r)/(1-r) \sim 1.6$, which means an increase, due to mixing,
of 60\% in the raw asymmetry of Eq.~\ref{eq17}. 

As in the previous case of $B^\pm$ production, similar formulas to those of
Eqs.~\ref{eq15} and \ref{eq17} can be obtained for the production asymmetry as a
function of $p_T^2$ and/or $x_F$.

%% file: cp-asymmetry.tex
\section{B-meson CP asymmetries $p-p$ collisions}

In flavor symmetric machines, like the Tevatron or $e^+-e^-$ colliders, it is
customary to measure CP asymmetries just by counting the number of particles and
anti-particles in a given decay mode, making use of
\begin{equation}
A_{CP} = \frac{N_{B^+\rightarrow f}-N_{B^-\rightarrow
    \bar{f}}}{N_{B^+\rightarrow f}+N_{B^-\rightarrow \bar{f}}}\;,
\label{eq3-1}
\end{equation}
where $B^+$ and $B^-$ are respectively the leading and no-leading particles
decaying into a given final state and its c.c. respectively. However, the use of
the above equation is incorrect in flavor asymmetric machines like the LHC
because $N_{B^+\rightarrow f}\pm N_{B^-\rightarrow \bar{f}}$ necessarily
contains production effects. In fact, it is rather straightforward to show that
the CP asymmetry, corrected by the effect of the production asymmetry, is given by
\begin{equation}
A_{CP}=\frac{\left[\frac{N_{B^+\rightarrow f} - N_{B^-\rightarrow
    \bar{f}}}{N_{B^+\rightarrow f} + N_{B^-\rightarrow
    \bar{f}}}\right] - A}{1 - A\left[\frac{N_{B^+\rightarrow f} -
      N_{B^-\rightarrow \bar{f}}}{N_{B^+\rightarrow f} + N_{B^-\rightarrow
    \bar{f}}}\right]}\;,
\label{eq3-1b}
\end{equation}
which is formally identical to Eq.~\ref{eq10} once the replacement $A
\leftrightarrow A_{CP}$ has been made. Note however that Eq.~\ref{eq3-1b} is of
no practical use unless the production asymmetry, $A$, is previously measured
independently of the CP asymmetry. The alternative is to use  
\begin{equation}
A_{CP}=\frac{\Gamma(B^+\rightarrow f) - \Gamma(B^-\rightarrow
  \bar{f})}{\Gamma(B^+\rightarrow f) + \Gamma(B^-\rightarrow \bar{f})}\;,
\label{eq3-2} 
\end{equation}
which is equivalent to Eq.~\ref{eq3} since $BR(B^\pm\rightarrow f) =
\Gamma(B^\pm\rightarrow f)/\tau_P$. The decay width $\Gamma$ is measured
independently of the production asymmetry by means of
\begin{equation}
\Gamma(B^\pm \rightarrow f) = -\frac{1}{N_{B^\pm \rightarrow f}}\frac{dN_{B^\pm
    \rightarrow f}}{dt} \;,
\label{eq3-3}
\end{equation}
while the lifetime $\tau_P$ can be measured making use of Eq.~\ref{eq3-3} and summing
over all the decay modes.

The above discussion is not merely academic, but of rather practical
consequences. In fact, there exist several CP related quantities which are
measured by determining $N_{B^+\rightarrow f}\pm N_{B^-\rightarrow
  \bar{f}}$. As an example let us consider a recently proposed method to search
for CP asymmetries in Dalitz analysis~\cite{bediaga}, which is based on the 
measurement, bin by bin in the Dalitz plot, of the quantity 
\begin{equation}
^{Dp}S_{CP}=\frac{N_{B^+\rightarrow f}(i)-N_{B^-\rightarrow
      \bar{f}}(i)}{\sqrt{N_{B^+\rightarrow f}(i)+N_{B^-\rightarrow
        \bar{f}}(i)}}\;.
\label{eq3-4}
\end{equation}
The quantity of Eq.~\ref{eq3-4} has the property of being Gaussian distributed
with mean $\mu = 0$ and width $\sigma=1$ in the limit of large number of events,
as can be easily seen by calculating the error in $^{Dp}S_{CP}$ as a function of
the errors in $N_{B^+\rightarrow f}(i)$ and $N_{B^-\rightarrow \bar{f}}(i)$ in
the limit of $N_{B^+\rightarrow f}(i) \rightarrow N_{B^-\rightarrow
  \bar{f}}(i)$, when no production and CP asymmetries are present. Effects due
to either production or CP asymmetries in $^{Dp}S_{CP}$ reveal through a shift
of the center and a modification of the witdh of the Gaussian.

In $p-p$ colliders where production effects are at work, the above quantity does
not measure the CP asymmetry alone, as discussed at the beginning of this
section, but a combined effect of both, the CP and production asymmetry. Thought
the production asymmetry is constant all over the Dalitz plot, the formula of
Eq.~\ref{eq3-4} has to be replaced by 
\begin{widetext}
\begin{equation}
^{Dp}S_{CP}^A=\sqrt{\frac{N_{B^+\rightarrow f}(i)+N_{B^-\rightarrow
      \bar{f}}(i)}{1-A}}\frac{\frac{N_{B^+\rightarrow f}(i)-N_{B^-\rightarrow
      \bar{f}}(i)}{N_{B^+\rightarrow f}(i)+N_{B^-\rightarrow
      \bar{f}}(i)}-A}{\sqrt{1-A\frac{N_{B^+\rightarrow f}(i)-N_{B^-\rightarrow
      \bar{f}}(i)}{N_{B^+\rightarrow f}(i)+N_{B^-\rightarrow
      \bar{f}}(i)}}}\;,
\label{eq3-5}
\end{equation}
\end{widetext}
to account for production effects. Eq.~\ref{eq3-5} can be obtained from
Eq.~\ref{eq3-4} with the replacement (A similar treatment is made in
  Ref.~\cite{astro}, in a different context.)  
\begin{equation}
N_{B^-\rightarrow
  \bar{f}}(i) \leftrightarrow N_{B^-\rightarrow
  \bar{f}}(i)\frac{N_{B^+}}{N_{B^-}} = N_{B^-\rightarrow  
  \bar{f}}(i)\frac{1+A}{1-A}\;,
\label{eq3-6}
\end{equation}
where $N_{B^+}(N_{B^-})$ is the total number of particles (antiparticles)
produced at the interaction vertex. In the absence of CP violating effects, 
$^{Dp}S_{CP}^A$ is Gaussian distributed with mean $\mu^A = 0$ and width
$\sigma^A$ which is a complicated function of $N_{B^+\rightarrow f}(i)$,
$N_{B^-\rightarrow \bar{f}}(i)$, $A$ and their errors. We refrain to
show the mathematical form of $\sigma^A$ because it is of no particular utility
and can be easily found by calculating the error on
$^{Dp}S_{CP}^A$. CP-violating effects can still be seen in $^{Dp}S_{CP}^A$ by
looking for deviations from $\mu^A=0$. In addition, it is important to remark
that the departure from zero of the mean $\mu^A$ and the value of $\sigma^A$
is dependent on the number of events in the particular bin/region of the Dalitz plot,
making absolute measurements of the effect impossible, unless the effect of the
statistics on $\mu^A$ and $\sigma^A$ be known. This behavior is due to the
presence of the square root in the denominator of Eqs.~\ref{eq3-4} and
\ref{eq3-5}. 

The use of the above method to look for CP violation in the Dalitz plot
presupposes that the production asymmetry $A$ has been measured independently of
the CP asymmetry, using a control channel free of CP-violating effects. In
addition, attention has to be paid to the fact that the asymmetry $A$ has to be
measured in the same momentum range in which the decaying particles are selected
for the Dalitz analysis. 

Of course, any detector effect leading to asymmetries in the measurement of the
number of particles and antiparticles has to be very well understood, since it
can be easily misinterpreted as a CP asymmetry.

%% file: conclusion.tex
\section{Conclusions and further discussion}

In this paper we have discussed the effect of CP violation and mixing in the
measurement of the production asymmetries. As shown in the text, the effect can
be non-negligible, of order of a few percent, even in cases in which the CP
symmetry is violated in the decay by a small amount. The effect of mixing is
much more pronounced.  

The CP asymmetries and mixing parameters can always be measured independently of
production effects in $p-p$ colliders, however paying attention to the fact that
any measurement has to be done properly, i.e. by resorting to their definitions
in terms of decay widths or branching fractions. In fact, in $p-\overline{p}$ or
$e^+-e^-$ colliders, where no production asymmetries are expected, quantities
like $N_{P\rightarrow f} \pm N_{\overline{P}\rightarrow \bar{f}}$ and
$\Gamma(P\rightarrow f) \pm \Gamma(\overline{P}\rightarrow \bar{f})$ are 
equivalent, however, in $p-p$ colliders they are not because the first acount for
both, differences in the decay widths and asymmetries in the production, while
the second has to do only with the dynamics of the decays. This is of particular
importance for the LHC experiments, in order to perform precision measurements
of CP asymmetries and mixing parameters.

For those CP-violating measurements relying on $N_{P\rightarrow f} \pm
N_{\overline{P}\rightarrow \bar{f}}$, as a general rule and to account for
production effects, $N_{\overline{P}\rightarrow \bar{f}}$ has to be {\it
  normalized} by a factor of $N_P/N_{\overline{P}}$, where $N_P$ and
$N_{\overline{P}}$ are respectively the total number of particles and
antiparticles produced at the interaction vertex, and measured in the same
momentum range in which the decaying particles have been reconstructed. That
normalization factor has to be determined by using a control channel free of
CP-violating effects. This is of particular importance in the determination of 
possible CP asymmetries in the charm sector since, if existent, they are
expected to be much smaller than any production effect. 

The reader should also
be aware that the quantity $^{Dp}S^A_{CP}~(^{Dp}S_{CP})$ is dependent on the
population of the bin, meaning that in different regions of the Dalitz plot, the
effect of possible CP asymmetries is different. The same is valid also for the
production asymmetry. In fact, the departure from $\mu=0$ of $^{Dp}S_{CP}$
depends on the population of the given region of the Dalitz plot, since numerator
and denominator of Eq.~\ref{eq3-4} scale in a different way with $N$.

Conversely, the measurement of particle/antiparticle production asymmetries
requires first of the determination of the CP asymmetries and/or mixing effects
in the particular decay channel in which particle and antiparticles are being
reconstructed. This cannot be avoided, as CP asymmetries and/or mixing effects
have to do with the dynamics of the particular decay mode in which particles are
reconstructed. Fortunately, CP asymmetries and mixing effects can always be
measured independently of production effects.

Finally, we would like to emphasize that, although along the text the discussion 
has been focused on the production of B-mesons in $p-p$ collisions, it can be
extended with almost no changes to the production of particles and antiparticles
in any flavor asymmetric machine.

%% file: paper.bbl
\begin{thebibliography}{99}
\bibitem{experiments} G.A. Alves {\it et al.} (E769 Collaboration),
  Phys. Rev. Lett. {\bf 69} (1992) 3147; M. Adamovich {\it et al.} (WA82
  Colaboration), Phys. Lett. {\bf B305} (1993) 402; G.A. Alves {\it et al.}
  (E769 Collaboration), Phys. Rev {\bf D49} (1994) R4317; {\it ibid.}
  Phys. Rev. Lett. {\bf 72} (1994) 812; E.M. Aitala {\it et al.} (E791
  Collaboration), Phys. Lett. {\bf B371} (1996) 157; M. Adamovich {\it et al.} 
  (WA92 Collaboration), Nucl Phys. {\bf B495} (1997) 3;  E.M. Aitala {\it et
    al.} (E791 Collaboration), Phys. Lett. {\bf B 495} (2000) 42; E.M. Aitala
  {\it et al.} (E791 Collaboration), Phys. Lett. {\bf B 496} (2000) 9;
  E.M. Aitala {\it et al.} (E791 Collaboration), Phys. Lett. {\bf B539} (2002)
  218. 

\bibitem{asymmetry} E. Cuautle, G. Herrera and J. Magnin, Eur. Phys. J. {\bf C
  2} (1998) 473; G. Herrera and J. Magnin, Eur. Phys. J. {\bf C 2} (1998) 477;
  J.C. Anjos, J. Magnin and G. Herrera, Phys. Lett. {\bf B 523} (2001) 29;
  G.H. Arakelian, A. Capella, A.B. Kaidalov, Yu.M.Shabelski, hep-ph/0103337;
  K. Boreskov, A. Capella, A. Kaidalov, J. Tran Thana Van, Phys. Rev. {\bf D47}
  (1993) 919; R. Vogt and S.J. Brodsky, {\it Nucl. Phys.} {\bf B478} (1996) 311.

\bibitem{intrinsic-sea} R. Vogt, S.J. Brodsky and P. Hoyer, Nucl. Phys. {\bf
  B383} (1992) 643; V. Barger, F. Halzen and W.Y. Keung, Phys. Rev. {\bf D24}
  (1981) 1428; {\it ibid.} Phys. Rev. {\bf D25} (1982) 112.

\bibitem{lhcb} J. Magnin {\it et al.} (LHCb Collaboration), PoS BEAUTY2009
  (2009) 026; P.M. Spradlin, ArXiv:0711.1661.

\bibitem{leader} E. Leader and E. Predazzi, {\bf ``An introduction to gauge
  theories and modern particle physics''}, Vol. 2, pag. 19.

\bibitem{selex} M. Kaya {\it et al.} (SELEX Collaboration),
  ArXiv:hep-ex/0302039.

\bibitem{pdg} K. Nakamura {\it et al.} (Particle Data Group),
  J. Phys. {\bf G37} (2010), 075021.

\bibitem{bediaga} I. Bediaga, I.I. Bigi, A. Gomes, G. Guerrer, J. Miranda and
  A.C. dos Reis, Phys. Rev. {\bf D80} (2009) 096006.

\bibitem{astro} Ti-Pei Li and Yu-Qiang Ma, Astrophys. J. {\bf 272} (1983) 317.

%\bibitem{decay} T. Aaltonen {\it et al.} (CDF Collaboration),
%  Phys. Rev. Lett. {\bf 100} (2008) 121803.

\end{thebibliography}
